\documentclass[11pt]{article}
\usepackage[utf8]{inputenc}
\usepackage[margin=1in]{geometry}
\usepackage{indentfirst}
\usepackage{fancyhdr}
\usepackage{epsfig,amsfonts,latexsym,amsmath,amssymb,bm,color,datenumber}
\usepackage{graphicx}
\usepackage{caption}
\usepackage{subcaption}
\usepackage{url}
\usepackage{graphicx}
\usepackage{xcolor}
\usepackage{dirtytalk}
\newcommand{\be}{\begin{equation}}
\newcommand{\ee}{\end{equation}}
\newcommand{\la}{\label}

\title{ The Influence of Mineral Texture on Fracture Geometry in Layered Geo-Architected Rock}
\author{Liyang Jiang$^1$, Hongkyu Yoon$^2$, Antonio Bobet$^3$, and Laura J. Pyrak-Nolte$^{1,3,4}$ }

\date{%
$^1$Department of Physics and Astronomy, Purdue University \\%
$^2$Geomechanics Department, Sandia National Laboratory \\%
$^3$Lyle School of Civil Engineering, Purdue University \\%
$^4$Department of Earth, Atmospheric and Planetary Sciences, Purdue University \\%
\today}

\usepackage{graphicx}
\begin{document}

\maketitle

\section{Abstract}
Rock is a complicated material because of the inherent heterogeneity in mineral phases and composition, even when extracted from the same rock mass.  The spatial variability in compositional and structural features prevent reproducible measurements of deformation, fracture formation and other physical and chemical properties.  Here, we use geo-architected  3D  printed  synthetic  gypsum rock to  show  that  mineral  texture  orientation  governs  the isotropy or anisotropy in fracture surface roughness and volumetric flow rate through tensile fractures. Failure load is governed by the orientation of depositional layering relative to loading direction but this effect can be hidden when an orientated  mineral texture is present within the layers.  We find that samples with certain mineral orientations are twice as strong as the weakest samples and that high values of sample strength correlate with rougher fracture surfaces. Mineral texture oriented parallel to the failure plane gives rise to anisotropic surface roughness and in turn to anisotropic flow rates. The  uniqueness  of  induced  fracture  roughness and the degree of anisotropy is  a  potential method for assessment of the orientation and relative bonding strengths of minerals in a rock. With this information, we will be able to predict isotropic or anisotropic flow rates through fractures which is vital to induced fracturing, geothermal energy production and CO$_2$ sequestration.

\section{Behavior of Anisotropic Rocks}
In nature, rock forms through different geological processes that generate compositional, textural, and structural features.  A primary structural feature of  sedimentary rocks is layering that arises from depositional, compactional and diagenetic processes.  In addition, within a layer, a rock also contains textural features related to the arrangement of mineral components that can range from interlocking crystals, to foliations, or to fragments with either preferred mineral orientation or amorphously distributed. Past research has shown that layers or preferred mineral orientations are known to affect the mechanical properties of rock leading to direction dependent or anisotropic elastic moduli.  As examples, the anisotropy in elastic properties of shale is attributed to preferred orientation of clay minerals  \cite{Jones1981,Vernik1992,Sayers2005,Wenk2007}, and the preferred orientation of minerals is assumed to explain the seismic anisotropy of the Earth's inner core \cite{Song1997}. 

\begin{figure}[ht]
\centering
\includegraphics [width=0.99\linewidth]{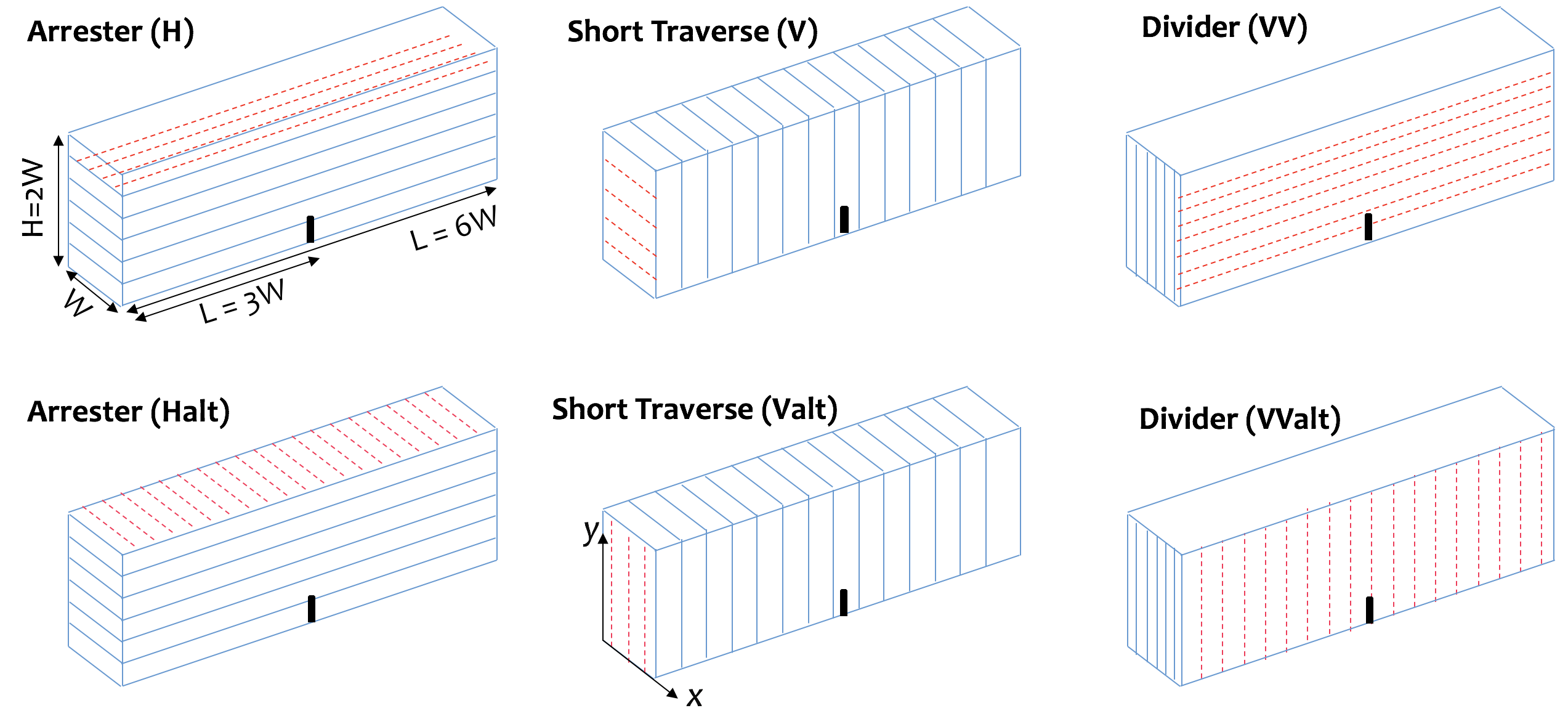}
\caption{Nomenclature of layered samples for tensile fractureing (a,b) Arrester, (c,d) Short Traverse, and (e,f) Divider.  Solid blue lines represent layering.  Red dashed lines represent binder direction during 3D printing.(Note: \textit{Not drawn to scale. For sample dimensions see Methods})} 
\label{fig:samples}
\end{figure}

A key question is how do compositional, textural, and structural features affect fracture formation and the resulting fracture geometry through which fluids flow.  Past and current research has shown that fracture toughness, i.e., the ability of a material to resist fracturing, is affected by layer orientation, with the geometry of the layers referred to as arrester, divider and short traverse (Figure \ref{fig:samples}).  For shale, many studies have observed that fracture toughness is ranked by the orientation of the layers with divider $>$ arrester $>$ short traverse \cite{Gao2017,Zheng2017}. However, other studies have observed cases where fracture toughness is comparable between arrester $\approx$  divider  or between arrester $\approx$ short traverse or values of fracture toughness for the short traverse specimens exhibit both the highest and lowest values \cite{Chandler2016}. These differences from the expected ranking of fracture toughness have been attributed to percent kerogen, inelasticity, clay, variable elastic properties among layers in shale, and also microfractures \cite{Schmidt1977,Chandler2016, ForbesInskip2018}.  

In this study, we examine the role of mineral texture orientation in layered rock on tensile fracture formation and geometry, and on fracture flow properties. Little is known as to whether mineral orientation within the layers affects tensile fracture formation and fracture toughness, especially when layering and minerals are not aligned.  To investigate the role of mineral orientation on tensile fracture formation and geometry, we performed three point bending experiments (3PB) on "geo-architected rock" with controlled directions of layering and mineral texture orientation to identify the contribution from each on tensile failure, surface roughness and permeability.


\section{Experiments on Geo-Architected Rock}

\begin{figure}[ht]
\centering
\includegraphics[width=0.99\linewidth]{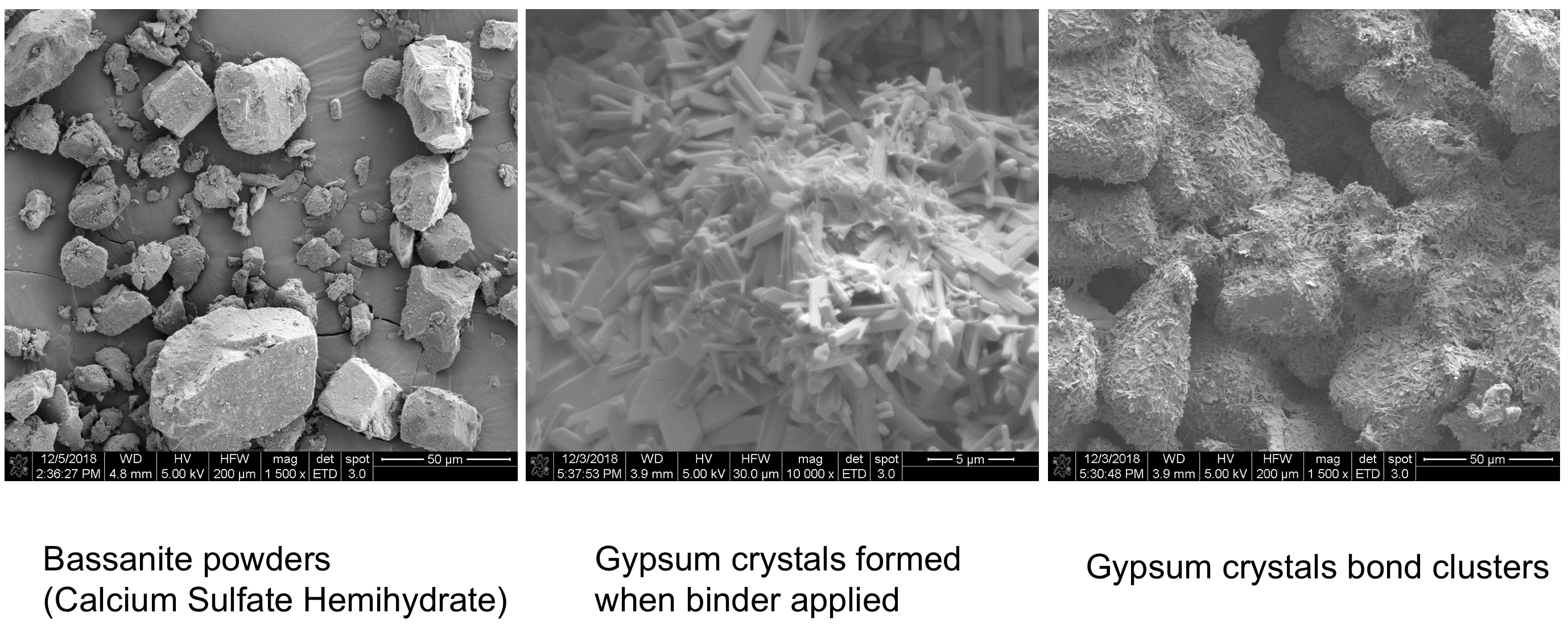}
\caption{Scanning Electron Microscopy (SEM) image of (left) bassanite powder;  (center) gypsum crystals formed from binder application, and (right) clusters of gypsum crystals.}
\label{fig:sem}
\end{figure}

Geo-architected rock refers to additively manufactured samples with repeatable physical and/or chemical properties to enable testing of hypotheses and model verification \cite{PyrakNolteDePaolo2015}.  For this study, geo-architected layered rock samples with preferred mineral fabrics were created using a 3D printing process (ProJet CJP 360).  Layers of calcium sulfate hemi-hydrate (0.1 mm thick bassanite powders, Figure \ref{fig:sem}) were bonded with a proprietary water-based binder (ProJet X60 VisiJet PXL) that produced gypsum as a reaction product (Figure \ref{fig:sem} center).  The gypsum mineral growth direction is oriented by the direction of the binder spreading.  When one layer of bassanite is deposited on a previous layer, gypsum crystals form a bond between bassanite layers after application of the binder.  Texture arises because the gypsum forms stronger bonds between gypsum crystals (Figure \ref{fig:sem} center) than between the gypsum crystals and bassanite powder.  Samples with different orientations of bassanite layers relative to gypsum mineral texture were printed to examine the effect of texture direction relative to layer direction on tensile fracture growth and the geometric properties of the induced fractures.

A three-point bending (3PB) test was used to induce a tensile fracture (Mode I) on printed specimens (25.4 x 76.2 x 12.7 mm$^3$) with a 5.08 mm long by 1.27 mm wide central notch to induce tensile failure (Figure ~\ref{fig:samples} where W = 25.4 mm).  For experiments performed with the in-situ stress rig in the 3D X-ray microscope, small 3PB samples were printed with a height of 4.8 mm, a length of 25 mm and a width of 4.2 mm and contained a central notch with a height of 0.96 mm.  Arrester (H \& Halt), short traverse (V \& Valt) and divider (VV \& VValt) samples were fabricated with different mineral orientations (red lines in Figure \ref{fig:samples}). Traditional cast gypsum samples with the same dimensions were used as a standard to compare to the behavior of the 3D printed gypsum samples. To achieve the same sample dimensions and notch size for the cast samples, a Formlabs 3D printer was used to create a acrylic-based sample from which a silicon rubber mold was formed.  The mold was filled with mixed gypsum and water and then vibrated to minimize the amount of trapped air. The cast gypsum samples were then cured in an oven at 40 \textsuperscript{o}C for 4 days. 

The geo-architected rock created through additive manufacturing exhibited anisotopic properties.  Anisotropy in 3D printed gypsum rock can rise from two sources: (1) the formation of bassanite layers during successive deposition of bassanite powder during manufacturing; and (2) the direction of mineral texture which is controlled by the binder application direction.  Ultrasonic measurements were performed to assess the isotropy/anisotropy of the geo-architected rock. From compressional ($P$) and shear wave ($S_H$ and $S_V$) velocity measurements, the samples were found to exhibit orthorhombic anisotropy (see Table \ref{speeds} in the Methods section). Unlike a transversely isotropic medium where the waves exhibit the same velocity when propagated parallel to the layering (i.e. from B-to-D and F-to-E in Figure \ref{fig:cube} in the Methods section), the geo-architected rock samples yielded $P$ and $S$ wave velocities that differed for these two propagation directions.  $P$, $S_H$ and $S_V$ exhibited the fastest velocities when propagated parallel to the direction of the mineral texture (red lines in Figure \ref{fig:cube}).  Interlocked gypsum crystals form preferentially along the line of binder application which enhances the mechanical properties in this direction.  This is similar to behavior observed in rock with preferred crystallographic  or shape orientation \cite{Wenk2007} and might potentially occur in igneous rock such as gabbro that exhibits obliquity between foliation and compositional layering  \cite{Benn1989}. 

3PB tests were performed to induce a tensile fracture in each specimen (see Methods for a description of 3PB).  Load and displacement measurements were recorded throughout the failure process (from prior to, during and post-peak failure) to examine the stiffness and brittleness/ductility of the samples.  After failure, spatial correlation lengths, asperity height probability distributions and microslope distributions were obtained from laser profilometry measurements of surface roughness of the induced tensile fracture.  Volumetric flow rates were estimated from numerical simulations using a flow network model \cite{PyrakNolteMorris2000,Petrovitch2013,Petrovitch2014,PyrakNolteNolte2016}.

\section{Influence of Mineral Texture on Failure Load}

\begin{figure}[ht]
\centering
	\includegraphics[width=0.85\linewidth]{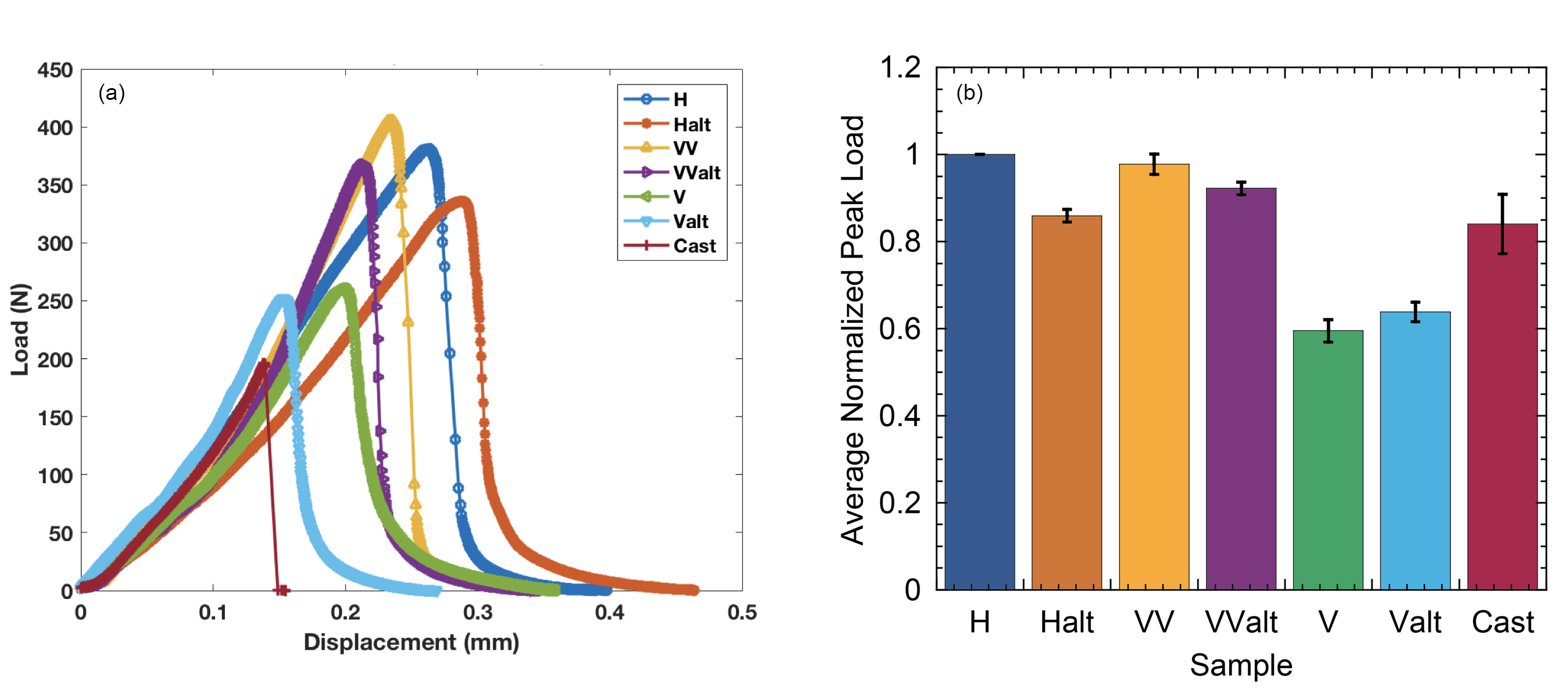}
\caption{ (a) Load-displacement graph for cast gypsum and representative 3D printed samples.  (b) Average relative peak load from 4 cohorts of samples for the cast and geo-architected samples. The samples are color-coded to match the colors in load-displacement curves (values are relative to H samples).}
\label{fig:loadrank}
\end{figure}

The co-existence of both layers and oriented mineral texture affects the interpretation of resistance to fracture that is simply based on layer orientation alone.  Load-displacement curves for the cast gypsum standard and the geo-architected rock are shown in Figure \ref{fig:loadrank}a.   From the load-displacement data, the peak failure load differed among the geo-architected samples even for samples with the same layer orientation.  For example, the peak failure load for arrester samples, H and Halt, differed though both contained layers that were oriented perpendicular to the loading.  H $>$ Halt because the fracture in H had to break across the gypsum crystals in order to propagate (Figure \ref{fig:samples}). The bonds between gypsum crystals (located between sequential bassanite layers) were stronger than the bonds between the gypsum and bassanite. The weakest geo-architected samples were the short traverse samples, V and Valt (Figure \ref{fig:loadrank}), both of which contained layers parallel to the direction of fracture propagation.      

To confirm that the observed differences in strength arise from mineral texture orientation and are not from variations in the samples, tests were conducted on multiple cohorts of samples. For cast samples, variation in material behavior can arise from impurities, pores and micro fractures \cite{Bobet1998} while variation in the 3D printed samples usually arises from equipment, printer settings and printer aging factors \cite{Kong2018}. To compare the sample strength for the different mineral texture orientations, 3PB tests were performed on samples cohorts that were printed together. The normalized average peak load (Figure \ref{fig:loadrank}b) was calculated based on data from four separate cohorts each containing the 6 tested geometries (Figure  1) for the 3D printed rocks.  For each cohort, the peak load was normalized by the peak load from arrester H, then the average of the 4 cohorts was taken. For cast samples, the average peak strength relative to the arrester H sample is based on data from 12 samples. These results indicate that layer orientation plays a dominant role in sample strength but layer orientation alone is not sufficient to predict relative resistance to fracturing in layered geologic materials.  Prediction requires knowledge of mineral texture orientation, especially in cases where layering and mineral texture may not be aligned. A key question is how the competing anisotropy between layering and mineral texture affects fracture propagation and surface roughness which strongly influences fluid flow through a fracture.


\section{Influence of Mineral Texture on Fracture Surface Roughness}

\begin{figure}[ht]
\centering
\includegraphics[width=0.90\linewidth]{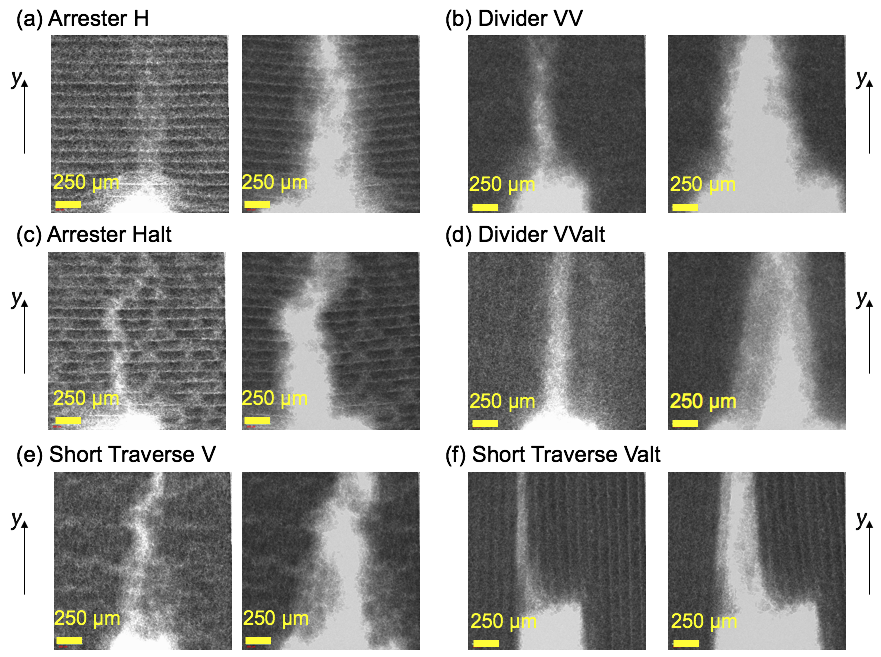}
\caption{2D x-ray radiographs of the small geo-architected samples at 5\% of peak load and just prior to complete failure.  The direction of fracture propagation from the notch (at the bottom of each image) is in the y-direction.  The x-direction is into the page.  } 
\label{fig:Deben}
\end{figure}

The factors that determine whether an induced tensile fracture is smooth or rough, or whether it exhibits direction-dependent roughness, depends on the relative resistance to failure among the rock constituents and structural features.  Both the layering and the mineral texture can cause a fracture to wander or deviate from a straight path, creating a roughness along the fracture surfaces.  For the induced tensile fractures in this study, the fracture propagation path was imaged using 3D X-ray microscopy and laser profilometry (see Methods).  The X-ray microscopy provides 2D radiographs and 3D reconstruction of specimens during and post-failure.  Comparison of the radiographs (Figure \ref{fig:Deben}) for the geo-architected samples indicates that, in general, the fracture trace is relatively straight when fractures are propagated parallel to the layering (e.g. Valt in Figure \ref{fig:Deben}f) and deviate from a straight path when propagating across layering (e.g.  H and Halt in Figures \ref{fig:Deben}a\&c).  However, the fracture trace for the short traverse sample V is not as straight as that observed for short traverse sample Valt even though they have the same layer orientation.  Therefore the difference in mineral orientation between these two samples affects the propagation path and indicates that mineral texture alters fracture propagation paths.  

\begin{figure}[ht]
\centering
\includegraphics[width=0.99\linewidth]{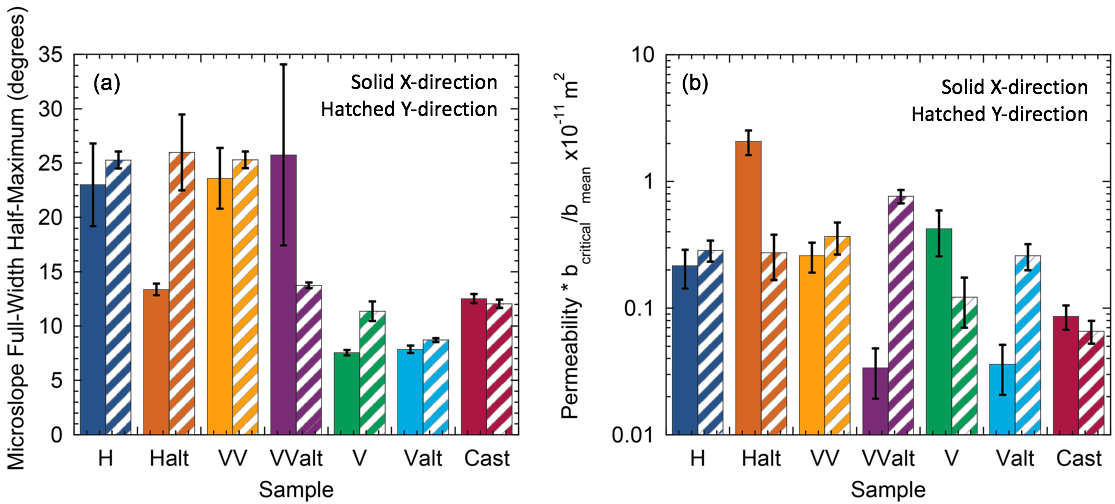}
\caption{(a) Average full width - half maximum of the microslope distribution for the cast gypsum and geo-architected samples which are color-coded to match the load-displacement curves in Figure \ref{fig:loadrank}. (b) Simulated fluid permeability based on surface roughness data from the cast gypsum and geo-architected samples (Figures \ref{fig:Gypsum_MS} \& \ref{fig:Roughness_AutoCorr}). Solid color: in the direction of fracture propagation (\textit{y-direction}); Hatched/shaded Color:  in the direction perpendicular to fracture propagation (\textit{x-direction})}
\label{fig:MS}
\end{figure}

\begin{figure}[ht]
\centering
\includegraphics[width=0.75\linewidth]{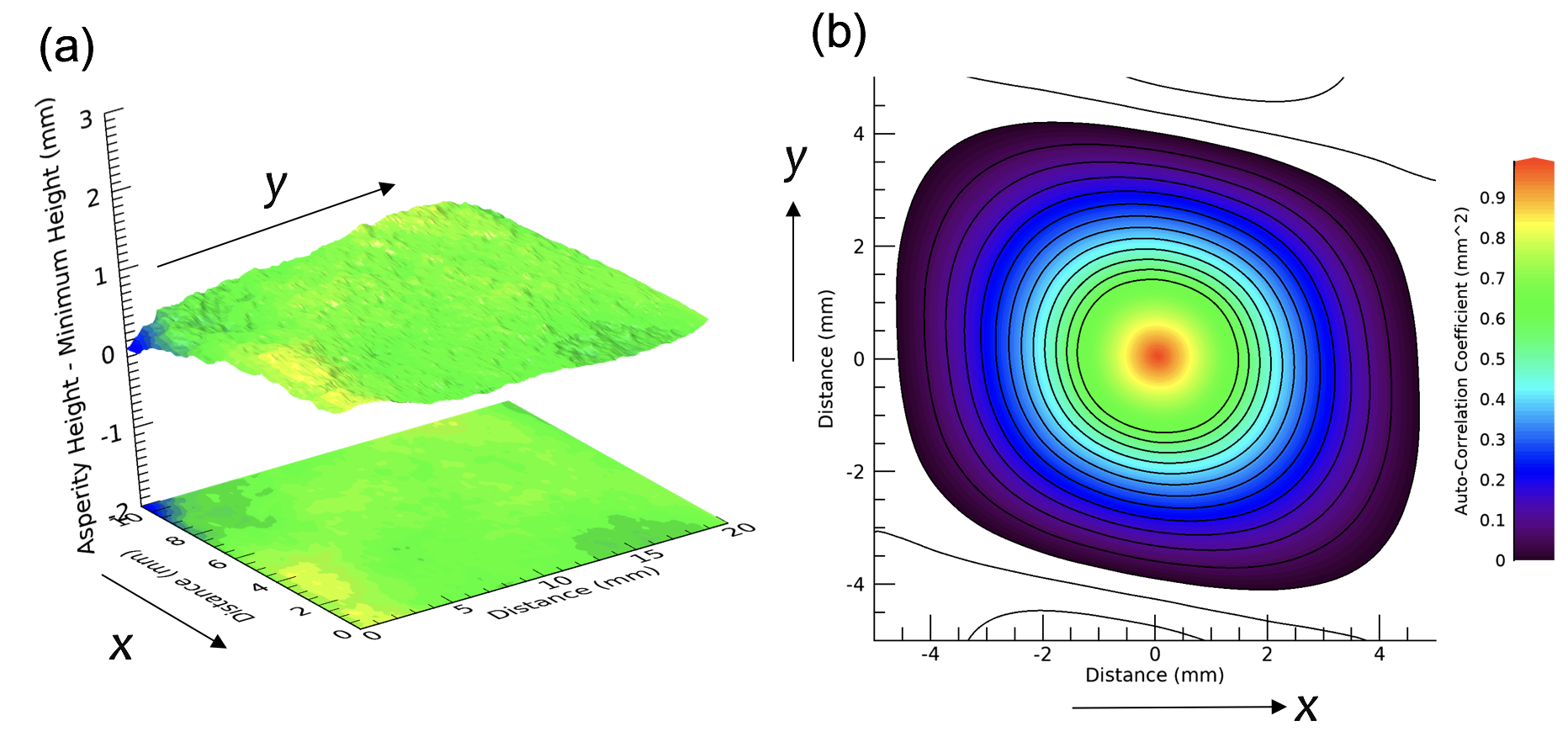}
\caption{(a) 3D surface and 2D contour of surface roughness for the cast gypsum sample (y direction is the direction of fracture propagation). (b) Normalized auto-correlation function for the cast gypsum sample.}
\label{fig:Gypsum_MS}
\end{figure}

\begin{figure}[ht]
\centering
\includegraphics[width=0.9\linewidth]{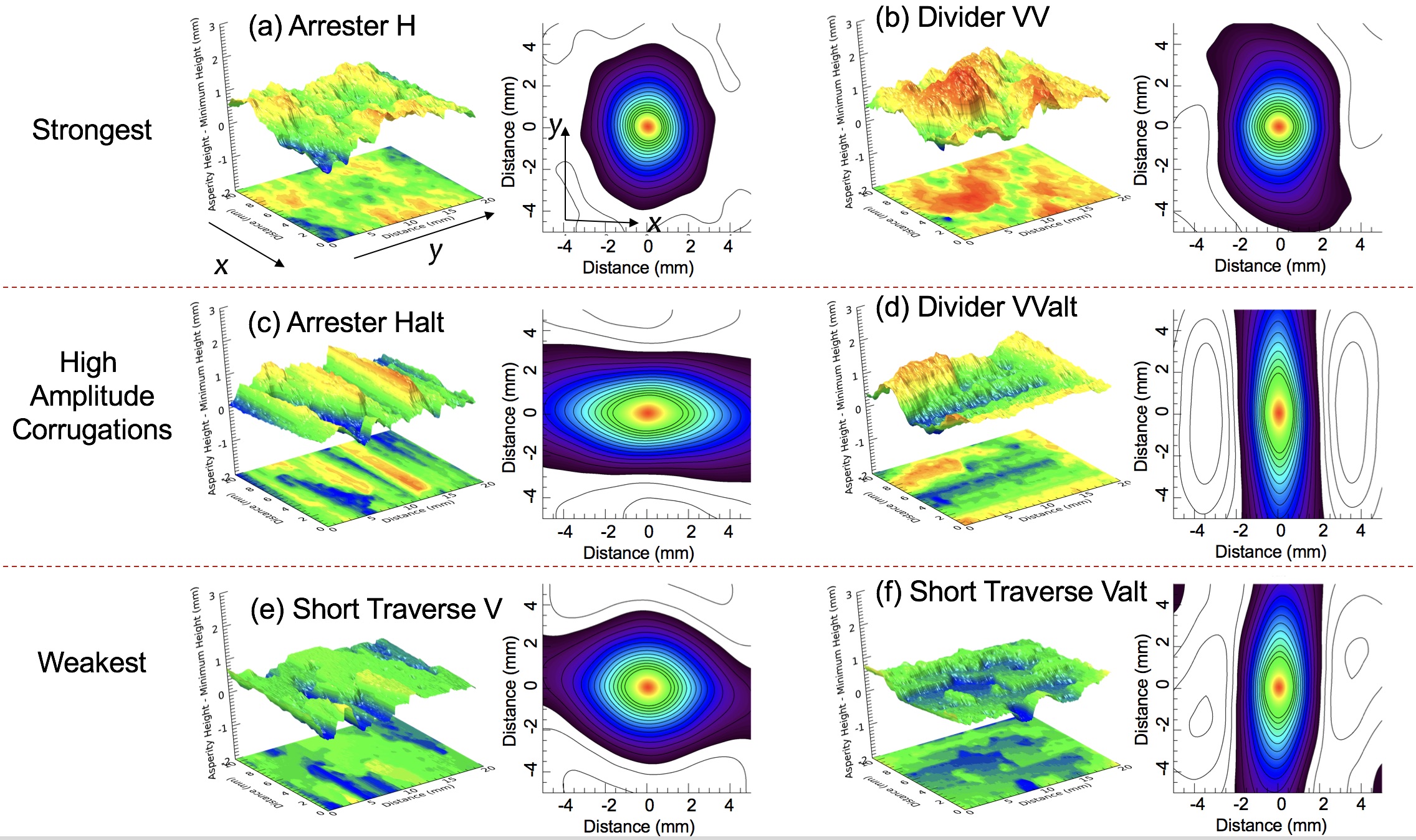}
\caption{3D surface and 2D contour of surface roughness and the normalized 2D autocorrelation function for the geo-architected samples grouped.  Note: All axes are the same as those shown in Figure \ref{fig:Gypsum_MS}. }
\label{fig:Roughness_AutoCorr}
\end{figure}

Asperity height measurements along the induced tensile fracture surfaces were made to quantify the effect of mineral texture orientation on surface roughness.  Microslope (Figure \ref{fig:MS}a) and autocorrelation analyses were applied to data acquired from laser profilometry  of the cast gypsum (Figure \ref{fig:Gypsum_MS}) and  geo-architected rock (Figure \ref{fig:Roughness_AutoCorr}).  Microslope analysis provides a method to quantify relative roughness (smooth: $\theta_{save} < 15^o$, rough $\theta_{save} > 15^o$, see Methods), and auto-correlation analysis provides a measure of isotropy or anisotropy in the spatial distribution of asperity heights. Details of the surface analysis approaches are given in the Methods section.  

From the microslope analysis (Figure \ref{fig:MS}a), when no layering and a uniform mineral distribution exists, such as in the cast gypsum sample, the surfaces of a tensile fracture are relatively smooth (Figure \ref{fig:Gypsum_MS}).  As expected for the short traverse samples (V and Valt), the surfaces were smooth ($\theta_{save} < 15^o$) because the fracture propagation path was parallel to the layering, i.e. breaking along the weaker bonds between the gypsum mineral and the bassanite layers.  The difference in the fracture trace observed for samples V and Valt in Figure \ref{fig:Deben} is evident in the microslope analysis where the surface is slightly rougher in the y-direction for V.  Direction dependent roughness is also observed for arrester sample Halt and divider sample VValt.  The smooth direction ($\theta_{save} < 15^o$) in both of these samples is parallel to the mineral texture orientation (red lines in Figure \ref{fig:samples}), while the rough direction ($\theta_{save} > 15^o$) occurs as the fracture propagates perpendicular to the mineral texture (y-direction for sample Halt, x-direction for sample Valt).  Samples H and VV are rough in both directions because the layers and mineral texture both provide resistance to fracturing but in orthogonal directions.  The mineral texture in sample H and VV results in additional roughness in the y- and x-directions, respectively, while the layering produces roughness in the x- and y-directions, respectively.  

\begin{figure}[ht]
\centering
\includegraphics[width=0.9\linewidth]{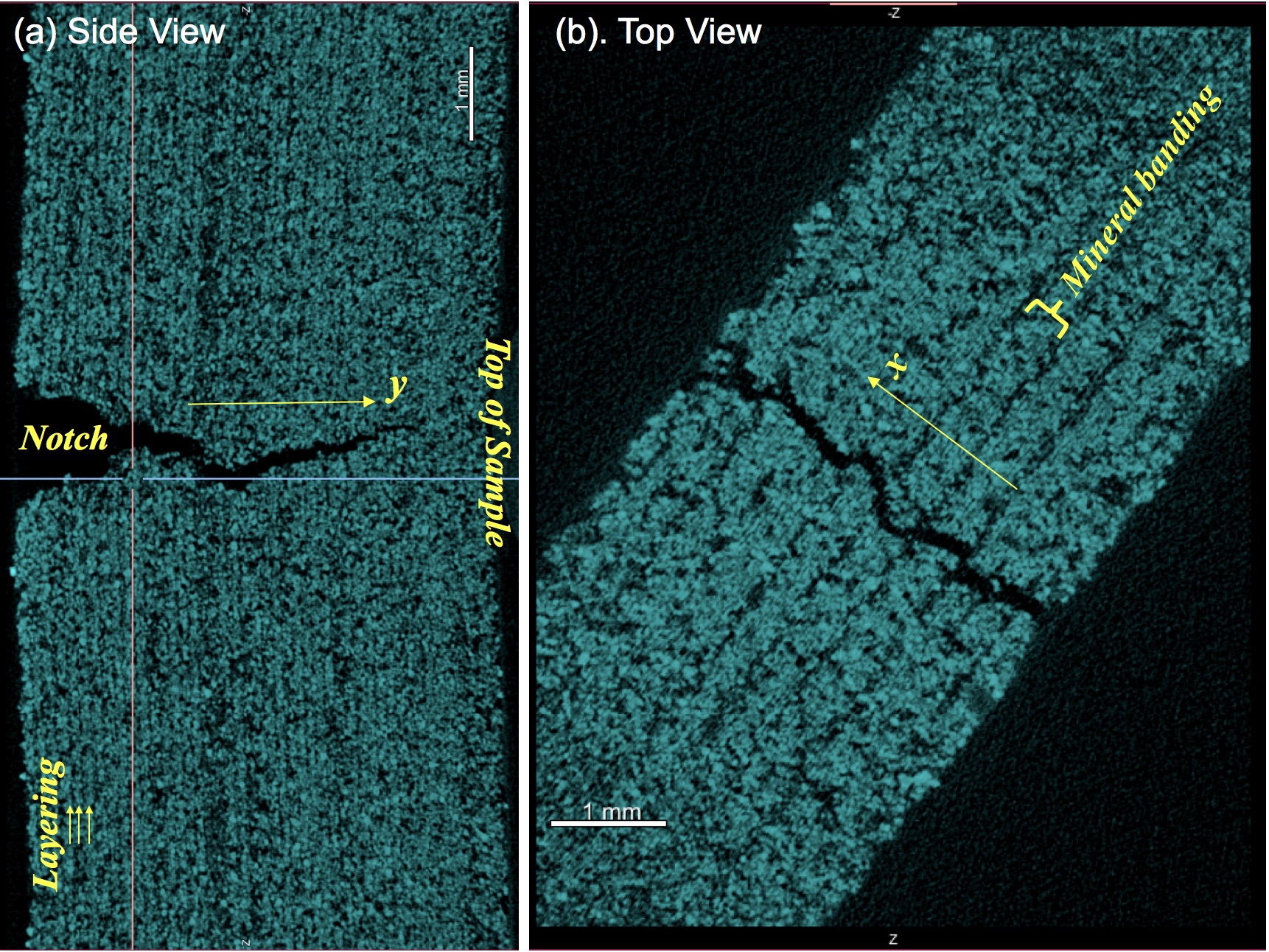}
\caption{Image from 3D X-ray tomographic reconstruction of small sample H post-peak showing the fracture trace in the (a) Side view showing the fracture trace in the direction of fracture propagation from notch to top of sample (y-direction) with layer direction indicated by the small yellow arrows. (b) Top view showing the fracture trace in the x-direction and gypsum mineral banding. Scale bars in each image represent 1 mm.}
\label{fig:H}
\end{figure}

These results indicate that there is an additional toughness that is associated with the difference in resistance to fracturing between the layering and the minerals.  3D X-ray microscopy was performed to examine this difference.  Figure \ref{fig:H} contains images from a 3D tomographic reconstruction of sample H after post-peak loading, showing the fracture trace in the y-direction (Figure \ref{fig:H}a) and the x-direction (Figure \ref{fig:H}b). In Figure \ref{fig:H}b, bands of mineral texture are observed, indicating the width of the spray from the binder application that results in the formation of gypsum crystals.  The fracture trace in the x-direction exhibits roughness on the scale of the width of the binder spray as the strength and amount of gypsum crystals is less between mineral bands.  In Figure \ref{fig:H}b, the fracture path is observed to wander around the mineral texture, seeking the path of least resistance through the mineral bands.   While in (Figure \ref{fig:H}a), the fracture is observed to deviate from a straight path because of the layering. This observation shows that the competition in resistance to fracturing between the layers and mineral texture affects the roughness of the induced tensile fractures.

Whether an induced tensile fracture exhibits isotropy or anisotropy in surface roughness (Figure \ref{fig:Roughness_AutoCorr}) also depends on both the layering and mineral texture directions relative to the direction of fracture propagation. The fracture surfaces from the cast gypsum sample exhibited isotropic asperity heights (Figure \ref{fig:Gypsum_MS}) because the mineral composition was homogeneous and no layers existed in the sample.    However, both isotropic and anisotropic surface roughness were observed for the geo-architected samples (Figure \ref{fig:Roughness_AutoCorr}).  While the weakest samples (short traverse V and Valt) were relatively smooth, the oriented mineral textures resulted in small amplitude corrugations that give rise to stronger spatial correlations in the direction of mineral texture (e.g. x-direction in V Figure \ref{fig:Roughness_AutoCorr}e; y-direction in Valt Figure \ref{fig:Roughness_AutoCorr}f).  For the strongest geo-architected samples (H and VV), the surfaces exhibited short-range isotropy as indicated by the nearly circular contour lines because the surfaces were rough both parallel and perpendicular to the direction of fracture propagation, leading to isotropic rough surfaces (Figure \ref{fig:Roughness_AutoCorr}a\&b). The strong anisotropy in surface roughness for samples Halt and Valt occurs from the high amplitude corrugations (Figure \ref{fig:Roughness_AutoCorr} c\&d) with valley/ridges in a preferred orientation.  These corrugations are on the scale of the mineral texture width (Figure \ref{fig:H}) and the ridges of the corrugation are aligned parallel to the direction of mineral texture (red lines in Figure \ref{fig:samples}).  The strong anisotropy in the asperity height distribution for Halt and VValt occurs because the layers and mineral texture both provide geometric toughening in the same direction, enhancing the roughness.

To summarize, when a propagating fracture does not cross the layers (e.g. V and Valt) the fracture surfaces tend to be smooth and the anisotropy is governed by the mineral texture orientation.  If a propagating fracture crosses the layers, the fracture surfaces tend to be rough (e.g. H, Halt, VV and VValt).  However, the isotropy/anisotropy of the roughness is controlled by the mineral texture orientation relative to the layering.  If the mineral texture and layer orientations are orthogonal (e.g. H and VV), the fracture surface roughness tends to be isotropic.  Conversely, if the mineral texture and layer orientation are aligned (e.g. Halt and VValt), strong anisotropy in surface roughness is observed (Figures \ref{fig:MS}a \& \ref{fig:Roughness_AutoCorr}c\&d). 

Several of the induced tensile fractures in the geo-architected samples had corrugated surfaces.  From this study, corrugation in surface roughness was suppressed in the strongest samples (H and V), but enhanced in the intermediate strength when the resistance to fracturing from both the mineral texture and layers were aligned.  The weakest samples resulted in the small-amplitude corrugation. Corrugated surfaces often exhibit preferred directions of fluid flow that are parallel to the ridges/valley.  Flow perpendicular to the corrugations is expected to be less because of the diferrences in path length.  

\section{Mineral Texture Controls on Fracture Fluid Flow}

Fluid flow through a fracture is intimately related to the roughness of the fracture surfaces and the flow path topology that is formed when the two surfaces are in contact \cite{Brown1987,Bout2006}.  Simulations of fluid flow were performed using the surface roughness measurements from each sample and the numerical method described in \cite{Petrovitch2014,PyrakNolteNolte2016}. The roughness from one fracture surface was placed in contact with a flat plane to simulate the flow path topology.  A contact area of 5\% was selected to reduce the arbitrariness in fracture aperture.  Figure \ref{fig:MS}b shows the simulated permeability for two orthogonal directions (parallel and perpendicular to the direction of fracture propagation) for the tensile fractures that formed in cast gypsum and the geo-architected samples.  The permeability is scaled by the ratio of the critical neck (smallest aperture along the dominant flow path) to the mean aperture to account for differences in the size of the apertures among the different samples and between orthogonal directions on the same sample.  With this normalization, any variations in flow rate are related to spatial correlations in the aperture distribution \cite{PyrakNolteNolte2016}.  

Samples with isotropic surface roughness (cast gypsum, H, VV) exhibited flow rates in the two orthogonal directions that were within 30\%.  For anisotropic surfaces (V, Valt, Halt and VValt), flow rates varied between the parallel and perpendicular directions of fracture propagation by factors of 4 to 40. For Halt and Valt, the samples exhibiting the strongest corrugations in surface roughness, the permeability is greater parallel to the ridges (y- and x- directions, respectively) than perpendicular to the ridges. The anisotropy observed from the microslope analysis (Figure \ref{fig:MS}a) manifests in permeability anisotropy for the fractures (Figure \ref{fig:MS}b).  This finding suggests that estimates of permeability isotropy or anisotropy could be potentially predicted from careful study of mineral texture orientation relative to layering and strength of each feature prior to fracturing a rock.

\section{Conclusions}

{Geo-architected rock is instrumental in unraveling the complexity and heterogeneity observed in fracture formation in natural rock.  Geo-architected rock enabled directional-control of mineral texture in layered samples in a repeatable manner.  Variability in peak strengths of the geo-architected rock was still observed ($\approx$10\%) but was less than that observed for natural rock ($\approx$25\%).  From this study, the degree of roughness of a fracture surface was dominated by the orientation of layering relative to the applied load.  But the isotropy or anisotropy in roughness was controlled by the relative orientation between the layers and mineral texture  along the failure path.  These results suggest that detailed mineralogical studies of mineral texture/fabric in laboratory samples is important to unravel failure strength, surface roughness, preferential flow paths, and how fractures propagate in layered geological media.

}
\section{Acknowledgment}
{Sandia National Laboratories is a multi-mission laboratory managed and operated by National Technology \& Engineering Solutions of Sandia, LLC, a wholly owned subsidiary of Honeywell International, Inc., for the U.S. Department of Energy's National Nuclear Security Administration under contract DE-NA0003525.  This work is supported by the Laboratory Directed Research and Development program at Sandia National Laboratories. This paper describes objective technical results and analysis. Any subjective views or opinions that might be expressed in the paper do not necessarily represent the views of the U.S. Department of Energy or the United States Government. The computational fluid analysis is based upon work supported by the U.S. Department of Energy, Office of Science, Office of Basic Energy Sciences, Geosciences Research Program under Award Number (DE-FG02-09ER16022). We also acknowledge the 3D X-Ray Microscope Facility in the Department of Physics for the images shown in this presentation, which were acquired on a Zeiss Xradia 510 Versa 3D X-ray Microscope that was supported by the EVPRP Major Multi-User Equipment Program 2017 at Purdue University.  
}
\clearpage
\section{Methods}
{
\subsection{Material Properties}
{Powder X-ray diffraction (XRD) was performed to determine the percent bassanite ($2Ca_2SO_4 \cdot H_2O$) and gypsum ($Ca_2SO_4 \cdot 2H_2O$) in the 3D printed samples.  The XRD system was a Panalytical Empyrean X-ray diffractometer equipped with Bragg-Brentano HD optics, a sealed tube copper X-ray source ($\lambda = 1.54178 \r{A}$), with soller slits on both the incident and receiving optics sides, and a PixCel3D Medipix detector. The anti-scatter slit (1/2$^o$) and divergence slit (1/8$^o$) as well as the mask (4 mm) were chosen based on sample area and starting $\theta$ angle. The XRD measurement found that the powder contained 97\% bassanite while the printed samples (i.e. after application of the binder) were $\approx$50-50 bassanite and gypsum.  The average density of the 3D printed geo-architected samples was $1190 \pm 5.5 kg/m^3$, while the average density for the cast gypsum samples was $1441 \pm 23.4 kg/m^3$.

Compressional and shear wave velocities were determined from ultrasonic waves measurements on a cube of cast gypsum and on a cube of 3D printed geo-architected gypsum samples. Olympus V103 (P-wave) and V153 (S-wave) piezoelectric transducers with a central frequency of 1 MHz were coupled to the sample with baked honey (8.75\% weight loss from the removal of water). An Olympus 5077PR pulse generator excited the source with 400V with 0.4 $\mu$s width with a 100Hz repetition rate. After propagating through the sample, the signals were digitized using a National Instruments USB-5133 digitizer and stored on a computer for analysis. A sampling rate of 100 MSamples/sec was used to get a bin size of 0.01 microseconds. The dimensions of the cubes and measured velocities are listed in Table \ref{speeds}.  

\begin{table}[ht]
\centering
\caption{Compressional ($P$) and Shear wave ($S$) Velocity measured on a cubic sample of 3D Printed Rock. Shear wave polarizations, $S_{H}$ and $S_{V}$, are shown in Figure \ref{fig:cube}.}
\label{speeds}
\begin{tabular}{|c|c|c|c|c|}
\hline
 Waves & Length (mm) & $P$ ($m/s$) & $S_{H}$ (m/s) & $S_{V}$ (m/s)  \\ \hline
 A-to-C & 49.7 & 2360  & 1475 & 1419  \\ \hline
 B-to-D & 49.7 & 2737  & 1549 & 1455   \\ \hline
 E-to-F & 49.3 &2430 & 1539  & 1381  \\ \hline
\end{tabular}
\end{table}

\begin{figure}[ht]
\centering
\includegraphics[width=2 in]{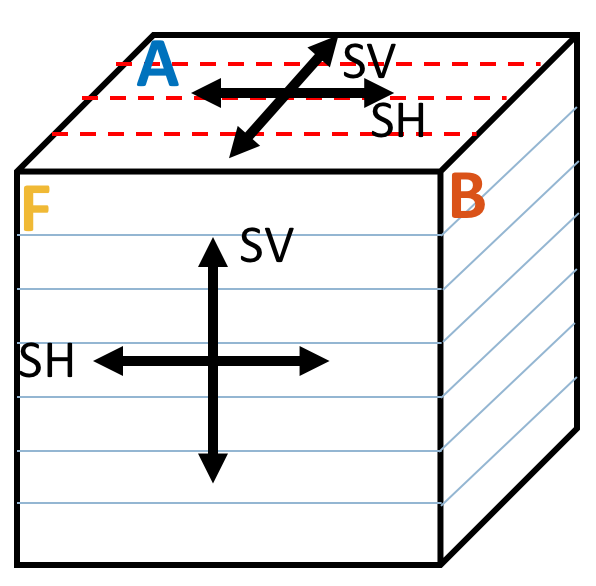}
\caption{Solid blue lines represent layering.  Red dashed lines represent direction of mineral fabric. (Note: \textit{Not drawn to scale.}) Shear wave polarizations are indicated by the black arrows.  Waves were propagated from A-to-C, B-to-D and E-to-F.}
\label{fig:cube}
\end{figure}

Unconfined compressive strength (UCS) testing was performed on cast gypsum and 3D printed gypsum samples with a diameter of 25.4 mm and a height of 50.8 mm using a ELE International Soil Testing uniaxial loading machine (capacity 8898 N) with a displacement rate of 0.03 mm/min.  Load and displacement data were recorded at a 5 Hz sampling rate.  Displacement and load were read to within 0.01 mm and 5 N.  Figure \ref{fig:UCS} shows the geometry of the cylinders used in the UCS testing and representative stress-strain curves for the cast gypsum and 3D printed samples. UCS values for cast gypsum and the 3D printed geo-architected samples are given in Table \ref{UCSValues}.

\begin{figure}[ht]
\centering
\includegraphics[scale=0.45]{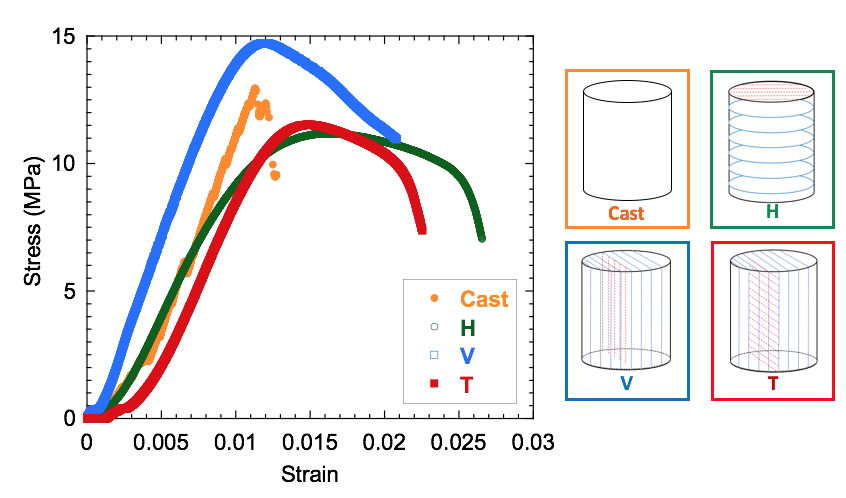}
\caption{Stress-strain curves for cast gypsum sample and 3D printed geo-architected samples H, V and T with layering (blue lines) and mineral texture (red lines) shown in the sketch of the samples. (Note: Layering and mineral texture are \textit{not drawn to scale.)}}
\label{fig:UCS}
\end{figure}

\begin{table}[]
\centering
\caption{Average Unconfined Compressive Strength (UCS) for the samples shown in Figure \ref{fig:UCS}.}
\label{UCSValues}
\begin{tabular}{|c|c|c|c|}
\hline
 Sample & UCS (MPa)  \\ \hline
 $Cast$ $Gypsum$ & $12.9$  \\ \hline
 $H$ & $11.2$   \\ \hline
 $T$ & $11.5$   \\ \hline
 $V$ & $14.7$   \\ \hline
\end{tabular}
\end{table}
}

\subsection{Three Point Bending Tests}
{
Tensile fractures (Mode I) were induced in samples using the three point bending test (e.g. \cite{Yu2018}). A rod was placed on the top surface of a sample directly aligned with the notch (Figure \ref{fig:samples}) and two rods were placed symmetrically on the bottom surface at a distance that was 10\% of the sample length (L in Figure 1) from the sides of the sample. Load was applied to a sample using an ELE International Soil Testing load frame with a 8896 N capacity S-shaped load cell. The loading rate was 0.03 mm/min.  Load and displacement (from a LVDT) data were recorded at a 5 Hz recording rate.
}
\subsection{Xray CT}
{
A 3D X-Ray Microscope (Zeiss Xradia 510 Versa) was used to acquire 2D radiographs and to perform 3D computed tomography for geo-architected geometry (Figure \ref{fig:samples}) during 3PB loading test.  The small samples were placed in a Deben CT5000 in-situ uniaxial loading device in the 3D X-ray microscope. The loading rate was  0.1 mm$/$min with load information recorded at 100 milisecond sampling rate.  The settings for both the 2D and 3D scans were 80 kV and 7W Xrays, at 4x magnification, with a source and detector distances of 70 mm and 200 mm, respectively, and 4s exposure time for each of the 1601 projections.  The voxel and pixel edge length (both for the 3D and 2D scans) was 1.75 $\mu$m.  The field of view was 1.78 mm$^2$.  Data reconstructions and anlaysis were performed using Object Research Systems (ORS) Dragonfly Pro 4.0 software.   
}
\subsection{Surface Roughness Analysis}
{Surface roughness measurements were made with a Keyence LK-G152 Laser with a wavelength of 650 nm.  The sample was mounted on coupled orthogonal translation stages (Newport MTM250PP1) controlled by a motion controller (Newport Universal Motion Controller ESP 300) to enable measurements of asperity height over a 2D area (105 mm by 200 mm) in increments of 0.1 mm.  The laser spot size was 120 $\mu m$.  First, the surface roughness maps were corrected from arbitrary rotations associated with mounting the sample in the laser profilometer system.  The gradients were determined by fitting a 2D plane to the surface and then subtracting the gradients from the asperity arrays.  Next, the minimum asperity height was subtracted from all points to yield asperity heights, $z_(x,y)$ that ranged from zero to the maximum for a given surface.  

The isotropy or anisotropy of a surface asperity height distribution was determined from a 2D auto-correlation analysis. In this approach, a 2D Fourier Transform, $FT$, of the asperity heights from a surface, $Z=FT(z(x,y))$, was multiplied by the complex conjugate, $\tilde{Z}$, and then an inverse Fourier transform, $FT^{-1}$, was performed on this product:

\begin{equation}
S(x,y)=\frac{FT^{-1}(Z*\tilde{Z})}{<z(x,y)*z(x,y)>}
\la{eq.06}
\end {equation}

and divided by the mean of the square of $z(x,y)$ to obtain $S(x,y)$, the 2D auto-correlation function.
The 2D asperity map was rectangular in shape which could bias or generate artifacts in $S(x,y)$.  For each surface roughness map, the auto-correlation function, $S(x,y)$, was calculated for 2 circular subregions (10 mm diameter). For each sample type (i.e. cast gypsum or 3D printed samples), the 2D auto-correlation functions shown in Figures  \ref{fig:Gypsum_MS} and \ref{fig:Roughness_AutoCorr} represent an average $<S(x,y)>$.  Figure \ref{fig:Gypsum_MS}a shows $S(x,y)$ for the cast gypsum sample.  The auto-correlation function indicates the probability that an asperity at a distance $r(x,y)$ will have a similar height.  The maximum probability is 1 when $r(x,y) = 0$ when a comparison is made between a point and itself.  The 2D images show that the roughness of the surface for the gypsum is isotropic, as indicated by the circular contours.  

Microslope angle analysis was also performed on the asperity height map, $z(x,y)$, to serve as a measure of the relative smoothness or roughness.  Park \& Song \cite{ParkSong2013} defined the microslope angle as the dip of the slope between neighboring asperities.  A microslope analysis was performed by finding the local slope, $s$, where

 \begin{equation}
 s_{x}=\frac{dz(x,y)}{dx}
\la{eq.11a}
\end{equation}
and
 \begin{equation}
s_{y}=\frac{dz(x,y)}{dy}
\la{eq.11b}
\end{equation}

which is the derivative of the surface roughness profile in the x-direction (horizontal and perpendicular to the direction of fracture propagation) and y-direction (vertical direction and parallel to the direction of fracture propagation).  The microslope angle is taken relative to the horizontal and is found by

 \begin{equation}
  \theta_{sx}=arctan(s_{x})
 \la{eq.12a}
 \end{equation}
  and 
\begin{equation}
  \theta_{sy}=arctan(s_{y}).
 \la{eq.12b}
\end{equation}

A surface was defined as relatively \say{smooth} if the average microslope angle ($\theta_{save}$) distribution full-width at half the maximum was $\theta_{save} < 15^o$. From the microslope analysis, one can see that the tensile fracture in gypsum was relatively smooth and had a narrow distribution centered at zero degrees with $\theta_{saveY} < 15^o$ in both the direction of fracture propagation (y) and perpendicular to fracture propagation (x).
}
}
\clearpage
\bibliography{references}

\begin{thebibliography}{10}
\expandafter\ifx\csname url\endcsname\relax
  \def\url#1{\texttt{#1}}\fi
\expandafter\ifx\csname urlprefix\endcsname\relax\def\urlprefix{URL }\fi
\providecommand{\bibinfo}[2]{#2}
\providecommand{\eprint}[2][]{\url{#2}}

\bibitem{Jones1981}
\bibinfo{author}{Jones, L. E.~A.} \& \bibinfo{author}{Wang, H.~F.}
\newblock \bibinfo{title}{Ultrasonic velocities in cretaceous shales from the
  williston basin}.
\newblock \emph{\bibinfo{journal}{Geophysics}} \textbf{\bibinfo{volume}{46}},
  \bibinfo{pages}{288--297} (\bibinfo{year}{1981}).

\bibitem{Vernik1992}
\bibinfo{author}{Vernik, L.} \& \bibinfo{author}{Nur, A.}
\newblock \bibinfo{title}{Ultrasonic velocity and anisotropy of hydrocarbon
  source rock}.
\newblock \emph{\bibinfo{journal}{Geophysics}} \textbf{\bibinfo{volume}{57}}
  (\bibinfo{year}{1992}).

\bibitem{Sayers2005}
\bibinfo{author}{Sayers, C.~M.}
\newblock \bibinfo{title}{Seismic anistropy of shales}.
\newblock \emph{\bibinfo{journal}{Geophysical Prospecting}}
  \textbf{\bibinfo{volume}{53}} (\bibinfo{year}{2005}).

\bibitem{Wenk2007}
\bibinfo{author}{Wenk, H.-R.}, \bibinfo{author}{Lonardelli, I.},
  \bibinfo{author}{Herman, F.}, \bibinfo{author}{Nihei, K.~T.} \&
  \bibinfo{author}{Nakagawa, S.}
\newblock \bibinfo{title}{Preferred orientation and elastic anisotropy of
  illite-rich shale}.
\newblock \emph{\bibinfo{journal}{Geophysics}} \textbf{\bibinfo{volume}{72}},
  \bibinfo{pages}{E69--E75} (\bibinfo{year}{2007}).

\bibitem{Song1997}
\bibinfo{author}{Song, X.}
\newblock \bibinfo{title}{Anisotropy of the earth’s inner core}.
\newblock \emph{\bibinfo{journal}{Reivews in Geophysics}}
  \textbf{\bibinfo{volume}{35}}, \bibinfo{pages}{297--313}
  (\bibinfo{year}{1997}).

\bibitem{Gao2017}
\bibinfo{author}{Gao, Y.} \emph{et~al.}
\newblock \bibinfo{title}{Theoretical and numerical prediction of crack path in
  material with anisotropic fracture toughness}.
\newblock \emph{\bibinfo{journal}{Engineering Fracture Mechanics}}
  \textbf{\bibinfo{volume}{180}}, \bibinfo{pages}{330--347}
  (\bibinfo{year}{2017}).

\bibitem{Zheng2017}
\bibinfo{author}{Zheng, X.} \& \bibinfo{author}{Wei, Y.}
\newblock \bibinfo{title}{Crack deflection in brittle media with heterogeneous
  interfaces and its application in shale fracking}.
\newblock \emph{\bibinfo{journal}{Journal of the Mechanics and Physics of
  Solids}} \textbf{\bibinfo{volume}{101}}, \bibinfo{pages}{235--249}
  (\bibinfo{year}{2017}).

\bibitem{Chandler2016}
\bibinfo{author}{Chandler, M.~R.}, \bibinfo{author}{Meredith, P.~G.},
  \bibinfo{author}{Brantut, N.} \& \bibinfo{author}{Crawford, B.~R.}
\newblock \bibinfo{title}{Fracture toughness anisotropy in shale}.
\newblock \emph{\bibinfo{journal}{Journal of Geophysical Research}}
  \textbf{\bibinfo{volume}{121}}, \bibinfo{pages}{1706--1729}
  (\bibinfo{year}{2016}).

\bibitem{Schmidt1977}
\bibinfo{author}{Schmidt, R.~A.}
\newblock \bibinfo{title}{Fracture mechanics of oil shale - unconfined fracture
  toughness, stress corrosion cracking, and tension test results}.
\newblock In \emph{\bibinfo{booktitle}{The 18th U.S. Symposium on Rock
  Mechanics}}, \bibinfo{pages}{2A21--2A26} (\bibinfo{publisher}{Balkema},
  \bibinfo{year}{1977}).

\bibitem{ForbesInskip2018}
\bibinfo{author}{Forbes~Inskip, N.~D.}, \bibinfo{author}{Meredith, P.~G.},
  \bibinfo{author}{Chandler, M.~R.} \& \bibinfo{author}{Gudmunsson, A.}
\newblock \bibinfo{title}{Fracture properties of nash point shale as a function
  of orientation to bedding}.
\newblock \emph{\bibinfo{journal}{Journal of Geophysical Research - Solid
  Earth}} \textbf{\bibinfo{volume}{123}}, \bibinfo{pages}{8428--8444}
  (\bibinfo{year}{2018}).

\bibitem{PyrakNolteDePaolo2015}
\bibinfo{author}{Pyrak-Nolte, L.} \& \bibinfo{author}{DePaolo, D.}
\newblock \bibinfo{title}{Controlling subsurface fractures and fluid flow: A
  basic research agenda}.
\newblock \bibinfo{type}{Report} (\bibinfo{year}{2015}).

\bibitem{Benn1989}
\bibinfo{author}{Benn, K.} \& \bibinfo{author}{Allard, B.}
\newblock \bibinfo{title}{Preferred mineral orientations related to magmatic
  flow in ophiolite layered gabrros}.
\newblock \emph{\bibinfo{journal}{Journal of Petrology}}
  \textbf{\bibinfo{volume}{30}}, \bibinfo{pages}{925--946}
  (\bibinfo{year}{1989}).

\bibitem{PyrakNolteMorris2000}
\bibinfo{author}{Pyrak-Nolte, L.~J.} \& \bibinfo{author}{Morris, J.~P.}
\newblock \bibinfo{title}{Single fractures under normal stress: The relation
  between fracture specific stiffness and fluid flow}.
\newblock \emph{\bibinfo{journal}{International Journal of Rock Mechanics and
  Mining Sciences}} \textbf{\bibinfo{volume}{37}}, \bibinfo{pages}{245--262}
  (\bibinfo{year}{2000}).

\bibitem{Petrovitch2013}
\bibinfo{author}{Petrovitch, C.}, \bibinfo{author}{Nolte, D.~D.} \&
  \bibinfo{author}{Pyrak-Nolte, L.}
\newblock \bibinfo{title}{Scaling of fluid flow versus fracture stiffness}.
\newblock \emph{\bibinfo{journal}{Geophysical Research Letters}}
  \textbf{\bibinfo{volume}{40}}, \bibinfo{pages}{2076--2080}
  (\bibinfo{year}{2013}).

\bibitem{Petrovitch2014}
\bibinfo{author}{Petrovitch, C.}, \bibinfo{author}{Pyrak-Nolte, L.} \&
  \bibinfo{author}{Nolte, D.~D.}
\newblock \bibinfo{title}{Combined scaling of fluid flow and seismic stiffness
  in single fractures}.
\newblock \emph{\bibinfo{journal}{Rock Mechanics And Rock Engineering}}
  (\bibinfo{year}{2014}).

\bibitem{PyrakNolteNolte2016}
\bibinfo{author}{Pyrak-Nolte, L.} \& \bibinfo{author}{Nolte, D.~D.}
\newblock \bibinfo{title}{Approaching a universal scaling relationship between
  fracture stiffness and fluid flow}.
\newblock \emph{\bibinfo{journal}{Nature Communications}}
  \textbf{\bibinfo{volume}{7}}, \bibinfo{pages}{Article Number 10663}
  (\bibinfo{year}{2016}).

\bibitem{Bobet1998}
\bibinfo{author}{Bobet, A.} \& \bibinfo{author}{Einstein, H.}
\newblock \bibinfo{title}{Fracture coalescence in rock-type materials under
  uniaxial and biaxial compression}.
\newblock \emph{\bibinfo{journal}{International Journal of Rock Mechanics and
  Mining Sciences}} \textbf{\bibinfo{volume}{35}}, \bibinfo{pages}{863--889}
  (\bibinfo{year}{1998}).

\bibitem{Kong2018}
\bibinfo{author}{Kong, L.}, \bibinfo{author}{Ostadhassan, M.},
  \bibinfo{author}{Li, M.} \& \bibinfo{author}{Tamimi, N.}
\newblock \bibinfo{title}{Can 3-d printed gypsum samples replicate natural
  rocks? an experimental study}.
\newblock \emph{\bibinfo{journal}{Rock Mech. Rock Eng.}}
  (\bibinfo{year}{2018}).

\bibitem{Brown1987}
\bibinfo{author}{Brown, S.~R.}
\newblock \bibinfo{title}{Fluid flow through rock joints: The effect of surface
  roughness}.
\newblock \emph{\bibinfo{journal}{Journal of Geophysical Research}}
  \textbf{\bibinfo{volume}{92}}, \bibinfo{pages}{1337--1347}
  (\bibinfo{year}{1987}).

\bibitem{Bout2006}
\bibinfo{author}{Boutt, D.~F.}, \bibinfo{author}{Grasselli, G.},
  \bibinfo{author}{Fredrich, J.~T.}, \bibinfo{author}{Cook, B.~K.} \&
  \bibinfo{author}{Williams, J.~R.}
\newblock \bibinfo{title}{Trapping zones: The effect of fracture roughness on
  the directional anisotropy of fluid flow and colloid transport in a single
  fracture}.
\newblock \emph{\bibinfo{journal}{Geophysical Research Letters}}
  \textbf{\bibinfo{volume}{33}} (\bibinfo{year}{2006}).

\bibitem{Yu2018}
\bibinfo{author}{Yu, M.}, \bibinfo{author}{Wei, C.}, \bibinfo{author}{Niu, L.},
  \bibinfo{author}{Li, S.} \& \bibinfo{author}{Yu, Y.}
\newblock \bibinfo{title}{Calculation for tensile strength and fracture
  toughness of granite with three kinds of grain sizes using
  three-point-bending tests}.
\newblock \emph{\bibinfo{journal}{PLoS ONE}} \textbf{\bibinfo{volume}{13}},
  \bibinfo{pages}{20} (\bibinfo{year}{2018}).

\bibitem{ParkSong2013}
\bibinfo{author}{Park, J.-W.} \& \bibinfo{author}{Song, J.-J.}
\newblock \bibinfo{title}{Numerical method for the determination of contact
  areas of a rock joint under normal and shear loads}.
\newblock \emph{\bibinfo{journal}{International Journal of Rock Mechanics and
  Mining Sciences}} \textbf{\bibinfo{volume}{58}}, \bibinfo{pages}{8--22}
  (\bibinfo{year}{2013}).

\end{thebibliography}
\end{document}